\documentclass[a4paper,11pt]{article}
\usepackage{epsf,multicol,ifthen}
\usepackage[cp1251]{inputenc}
\usepackage[dvips]{graphicx}
\usepackage{cite}
\usepackage{hyperref}
\title{ The realization of higher-dimensional breaking mechanism}

\author{T. Obikhod\thanks{E-mail: obikhod@kinr.kiev.ua}\\
\small\emph{Institute for Nuclear Research, National Academy of Science of Ukraine} \\
\small\emph{47, prosp. Nauki, Kiev, 03680, Ukraine}}

\date{\small\today}

\begin{document}

\maketitle

\abstract{We study D-branes on Calabi-Yau threefolds, which are realized through the blowing up the singularity of orbifold. This D-branes are represented as sheaves, which can be stable or unstable, what is connected with the transition in the Teichm$\ddot{u}$ller space. Using the derived category of McKay quiver representations, which describe D-branes as quivers and open superstrings between them by Ext groups, we can represent Higgs multiplets by the moduli space of an open superstring, connecting two McKay quivers. Through the equivalence between the derived category of coherent sheaves and triangulated category of distinguished triangles over the abelian category of McKay quivers we can associate D-branes with quivers or with sheaves, defined on Calabi-Yau. After the dimensional reduction of the ten-dimensional space-time we can receive matter content of the four-dimensional space-time. Thus, a higher-dimensional breaking mechanism is associated with four-dimensional GUT Higgs multiplets and symmetry breaking higgs mechanism. }

\section{Introduction}
\label{sec:intro}

The definition of D-branes as allowed endpoints for open strings \cite{1.}, generalizes the notion of quarks on which the QCD string can terminate. In contrast to the quarks of QCD, D-branes are intrinsic excitations of the fundamental theory. D-particles can probe distances much smaller than the size of the fundamental string quanta. D-branes played a crucial role in the 'second string revolution' , the way to reconcile quantum mechanics and Einstein gravity. The D-brane concept \cite{2., 3.} is powerfull because of the relations between supersymmetric gauge theories and geometry. 
	
	The purpose of our article is connected with the searches of a higher-dimensional breaking mechanism in the context of D-branes, which is connected or associated with four-dimensional Grand Unification Theory Higgs multiplets and symmetry breaking higgs mechanism. 

\section{The Category of D-branes as Derived Category of Coherent Sheaves}
	
	Due to the important development in string theory through the discovery of D-branes we can use a compactification model: the string theory has a target space $R^{1, 3}\times X$ for compact space $X$ and focus on $X$. Let $X$ be a topological space. On such space we can construct locally free sheaf ${\cal {E}}$. If we have embedding $i:S\rightarrow X$  and a sheaf ${\cal {E}}$ on $S$, than we can define a sheaf $i_{*}{\cal {E}}$ on $X$, through the following construction: a map $f:X\rightarrow Y$  between two algebraic varieties and a sheaf ${\cal {F}}$ on $X$ define the sheaf  $f_{*}{\cal {F}}$ on $Y$ by $f_{*}{\cal {F}}(U)={\cal {F}}(f^{-1}U)$. For embedding $i:X\rightarrow P^{n}$ the sheaf $i_{*}{\cal {E}}$ is given by $i_{*}{\cal {E}}(U)={\cal {E}}(U\cap P^{n})$ for all open subsets $U\subset P^{n}$. This embeds the sheaves on $X$ into the sheaves on $P^{n}$. So we have locally free sheaf on $P^{n}$, which is associated with the D-brane.  
	
	From \cite{4.} it follows that more generally we can consider  a complex of locally-free sheaves:
\[\ldots\stackrel{d_{n-1}}\longrightarrow{\cal{E}}^n\stackrel{d_{n}}\longrightarrow{\cal{E}}^{n+1}\stackrel{d_{n+1}}\longrightarrow{\cal{E}}^{n+2}\stackrel{d_{n+2}}\longrightarrow\ldots  ,\]
where morphisms $d_n:{\cal{E}}^{n}\rightarrow{\cal{E}}^{n+1}$, 
$d_n\in\mbox{Ext}^0({\cal{E}}^{n},{\cal{E}}^{n+1})=\mbox{Hom}({\cal{E}}^{n},{\cal{E}}^{n+1})$, $d_{n+1}d_{n}=0$ for all $n$ are morphisms between locally-free sheaves ${\cal{E}}^{n}$ and ${\cal{E}}^{n+1}$.\\
For a  further aim we must use the notion of a category.

	A category ${\cal{L}}$ consists of the following data:

	1) A class $\mbox{Ob}\ {\cal{L}}$ of objects $A, B, C, \cdots$;

	2) A\hspace*{2mm} family\ of\hspace*{2mm} disjoint\hspace*{2mm} 
sets\hspace*{2mm} 
of\hspace*{2mm} morphisms\ $\mbox{Hom}(A, B)$
one for each ordered pair $A, B$ of objects;

	3) A family of maps
\[\mbox{Hom}(A, B)\times \mbox{Hom}(B, C)\rightarrow \mbox{Hom}(A, C)\ ,\]
one for each ordered triplet $A, B, C$ of objects.

	These data obey the axioms:

	a) If $f\ : A\rightarrow B,\ g\ : B\rightarrow C,\ h\ : C\rightarrow D$,
then composition of morphisms is associative, that is, $h(gf)=(hg)f\ ;$

	b) To each object $B$ there exists a morphism $1_B\ :\ B\rightarrow B$
such that $1_Bf=f\ ,\ g1_B=g$ for $f\ : A \rightarrow B$ and 
$g\ : B \rightarrow C$\  .

	Thus, the category of D-branes is the
derived category of locally-free sheaves.
Locally-free sheaves and morphisms don't form an abelian category.
So we should replace the category of locally-free sheaves
by the abelian category of coherent sheaves. Abelian category is characterized by the existence of an exact sequences. 
We can form the category of D-branes, which is the derived category of coherent sheaves $D(X)$.
On a smooth space $X$, any coherent sheaf ${\cal {A}}$ has a locally-free resolution
\[0\rightarrow{\cal {F}}^{-3}\rightarrow{\cal {F}}^{-2}
\rightarrow{\cal {F}}^{-1}\rightarrow{\cal {F}}^{0}\rightarrow{\cal {A}}\rightarrow 0 \ , \]
where ${\cal {F}}^{k}$ is locally free. This is a quasi-isomorphism
${\cal {F}}^{\bullet}\rightarrow{\cal {A}}$ between
a complex of locally-free sheaves and a coherent sheaf.
Thus, an abelian category of coherent sheaves $D(X)$ of $X$
consists of objects ${\cal{E}}^{\bullet}$ - exact complexes of sheaves:
\[\ldots\stackrel{d_{n=2}}\longrightarrow{\cal {E}}^{n-1}\stackrel{d_{n=1}}\longrightarrow{\cal {E}}^{n}\stackrel{d_{n}}\longrightarrow
{\cal {E}}^{n+1}\stackrel{d_{n+1}}\longrightarrow \ldots \ \]
and morphisms between them ${\cal{E}}^{\bullet}\rightarrow{\cal{F}}^{\bullet}$:
\[\begin{tabular}{ccccccccc}
$\ldots$&$\stackrel{d_{n=2}}\longrightarrow$&${\cal {E}}^{n-1}$&$\stackrel{d_{n=1}}\longrightarrow$&${\cal {E}}^{n}$&$\stackrel{d_{n}}\longrightarrow$&${\cal {E}}^{n+1}$&$\stackrel{d_{n+1}}\longrightarrow$&$ \ldots$\\
&&&&&&&&\\
&&$\downarrow$&&$\downarrow$&&$\downarrow$&&\\
&&&&&&&&\\
$\ldots$&$\stackrel{d_{n=2}}\longrightarrow$&${\cal {F}}^{n-1}$&$\stackrel{d_{n=1}}\longrightarrow$&${\cal {F}}^{n}$&$\stackrel{d_{n}}\longrightarrow$&${\cal {F}}^{n+1}$&$\stackrel{d_{n+1}}\longrightarrow$&$ \ldots$

\end{tabular} \]

\section{Triangulated Category and Central Charge}

	For physical purposes we will work in future with Calabi-Yau threefolds $X$. According to \cite{4.}, if $X$ and $Y$ are mirror Calabi-Yau threefolds then the derived category $D(X)$ is equivalent to the triangulated category Tr${\cal{F}}(Y)$. So, in such category $D(X)$ objects are distinguished triangles:
\begin{center}
\hspace*{4cm}\begin{tabular}{ccc}
&$C$&\\
&\hspace*{-4mm}${\mbox{\tiny[1]}}\hspace*{-1mm}\swarrow$\hspace*{1mm}$\nwarrow$& \hspace*{15mm}$C={\mbox{Cone}}(f)$ \hspace*{42mm} (1)\\
&$\hspace*{-2mm}A$\hspace*{1mm}$\stackrel{f}{\longrightarrow}$\hspace*{1mm}$B$&\\
\end{tabular}	
\end{center}
and morphisms of this category are morphisms of distinguished triangles \cite{4.}.
Than the derived category is additive category, where exact sequences are exchanged by distinguished triangles.
According to Douglas \cite{5., 6.} instead of physical D-branes, living in the boundary conformal field theory, we can work with topological D-branes, and the
relationship between them is the notion of $\Pi$-stability.
A topological D-brane is physical if it is $\Pi$-stable. For the precise
definition of $\Pi$-stability we must compute the central charge of objects
${\cal{E}}^{\bullet}$ in $D(X)$:
\[Z({\cal{E}}^{\bullet})=\int\limits_X e^{-(B+iJ)}ch({\cal{E}}^{\bullet})
\sqrt{td(X)},\]
where $B+iJ$ is the complexified K$\ddot{a}$hler form.
For a given ${\cal{E}}^{\bullet}$ we may choose a grading 
$\xi({\cal{E}}^{\bullet})$:
\[\xi({\cal{E}}^{\bullet})=\frac{1}{\pi}\mbox{arg}Z({\cal{E}}^{\bullet}) \ \ \mbox{(mod 2)}.\]
If we have a distinguished triangle in $D(X)$ of the form (1)
with $A$ and $B$ stable, than $C$ is stable with respect to the decay represented by this triangle if and only if 
$\xi(B)\prec \xi(A)+1$. Also, if $\xi(B)= \xi(A)+1$ then 
$C$ is marginally stable and we may state that
\[\xi(C)=\xi(B)= \xi(A)+1 .\]
We may generalize this to the case of decays into any 
number of objects. For any object $E$ we may
define the set of distinguished triangles
\begin{center}
\hspace*{5cm}\begin{tabular}{ccccc}
&\hspace*{-2mm}$E_{0}$ \ $\longrightarrow$ \ \ $E_{1}$&
\hspace*{-2mm}$\longrightarrow$\ \ \ $\ldots$&
$\longrightarrow$ $E_{n}$ \  $=E$&\\
&\hspace*{-4mm}${\mbox{\tiny[1]}}$\hspace*{-1mm}$\nwarrow$\hspace*{1mm}$\swarrow$&
\hspace*{-8mm}${\mbox{\tiny[1]}}$\hspace*{-1mm}$\nwarrow$\hspace*{1mm}$\swarrow$&
\hspace*{-20mm}${\mbox{\tiny[1]}}$\hspace*{-1mm}$\nwarrow$\hspace*{1mm}$\swarrow$& \hspace*{2.3cm} (2)
 \\
&$A_{1}$&\hspace*{-5mm}$A_{2}$&\hspace*{-20mm}$A_{n}$&\\
\end{tabular}	
\end{center}
Then $E$ decays into $A_{1}, A_{2},\ldots A_{n}$ 
so long as
\[\xi(A_{1})\succ \xi(A_{2})\succ \ldots \xi(A_{n}).\]
As we have pointed out, $X$ and $Y$ are mirror Calabi-Yau 
threefolds. According to \cite{7.} mirror map is defined 
between the complex moduli space of a Calabi-Yau 
manifold $X$ and the K$\ddot{a}$hler moduli space of its 
mirror manifold $Y$. So we can work with K$\ddot{a}$hler moduli space, that is characterized by the complexified K$\ddot{a}$hler form $B+iJ$. Then we can use the Teichm$\ddot{u}$ller space
${\cal{T}}$ as the universal cover of the moduli space of
$B+iJ$ and we expect the set of stable D-branes to be well-defined
at any point in ${\cal{T}}$. \\
The following rules are applied:\\
$\bullet$ We begin with a stable set of D-branes with value of
grading $\xi$ for each D-brane.\\
$\bullet$ During the moving along a path in moduli space the
gradings will change continuously.\\
$\bullet$ Two stable D-branes may bind to form a new stable state.\\
$\bullet$ A stable D-brane may decay into other stable states.\\
So, we can write {\bf Conjecture} from \cite{4.}.\\
{\bf Conjecture.} At every point in the Teichm$\ddot{u}$ller space of $B+iJ$ there is a set of stable objects in $D(X)$ such that 
every object $E$ can be written in the form (2) for some $n$ and
for stable objects $A_{k}$.
 
	Every object in $D(X)$ is stable or unstable for a given
point in the Teichm$\ddot{u}$ller space of $B+iJ$. Thus the 
open string corresponding to $f$ in (1) go from tachyonic to massive
as we pass in the Teichm$\ddot{u}$ller space.

	Now as we work in ten-dimensional space-time $R^{3+1}\times {\newfont{\blackboard}{msbm8 scaled\magstep2}
\newcommand{\Ce}{\mbox{\blackboard\symbol{'103}}}\Ce}/G$
and knowing that after blowing up the singularity of additional six-dimensional space-time - orbifold 
${\newfont{\blackboard}{msbm8 scaled\magstep2}
\newcommand{\Ce}{\mbox{\blackboard\symbol{'103}}}\Ce}/Z_3 \rightarrow
{\cal{O}}_{P_{2}}(-3)$ we have the sheaf ${\cal{O}}_{P_{2}}(-3)$
on two-dimensional projective space $P_{2}$.  Here we must consider {\bf Theorem}  from \cite{4.}:\\
{\bf Theorem}. Suppose $X$ is a smooth resolution of the orbifold
${\newfont{\blackboard}{msbm8 scaled\magstep2}
\newcommand{\Ce}{\mbox{\blackboard\symbol{'103}}}\Ce}/G$ with 
$G$ a finite subgroup of $SU(d)$ and $d\leq 3$. Then the derived
category $D(X)$ is equivalent to the derived category of 
$G$-equivariant sheaves on ${\newfont{\blackboard}{msbm8 scaled\magstep2}
\newcommand{\Ce}{\mbox{\blackboard\symbol{'103}}}\Ce}^d$.\\
Now we are dealing with the derived category of sheaves on $P_2$ and
we can use the statement \cite{4.}:
D-branes on the orbifold 
${\newfont{\blackboard}{msbm8 scaled\magstep2}
\newcommand{\Ce}{\mbox{\blackboard\symbol{'103}}}\Ce}/G$
and open strings between them are described by the 
derived category of McKay quiver representations.\\
In future we will work with the derived category
of distinguished triangles over the abelian category
of McKay quivers. Objects of this category
are distinguished triangles (Figure 1)
\begin{figure}[htbp]
\centering{\includegraphics[width=.45\textwidth]{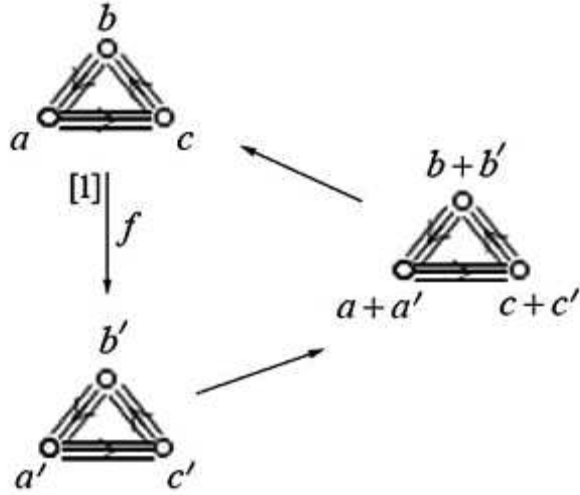}}
\caption{\label{fig:1} \small Construction of distinguished triangle.}
\end{figure}

(numbers $a, b, c$ and 
$a^{\prime}, b^{\prime}, c^{\prime}$ denote orbifold charges  characterizing McKay quivers); morphisms of this category are morphisms of 
distinguished triangles.

\section{Grand Unification Theory Breaking and Dimensional Reduction}

	Our further work will be connected with the efforts to take many attractive features of the basic Grand Unification Theory and implement this ideas in four-dimensional models, for example, in the minimal four-dimensional supersymmetric SU(5) GUT with standard Higgs content. Moreover, because no appropriate four-dimensional GUT Higgs field is typically available to break the GUT group to the Standard Model gauge group, it is necessary to employ a higher-dimensional breaking mechanism. For type IIB theories, the corresponding vacua are realized as compactifications of F-theory on Calabi-Yau fourfolds. We will consider the left part of Figure 2 of the general overview of how GUT breaking constrains the type of GUT model \cite{7.}.
\begin{figure}[htbp]
\centering 
\includegraphics[width=.45\textwidth]{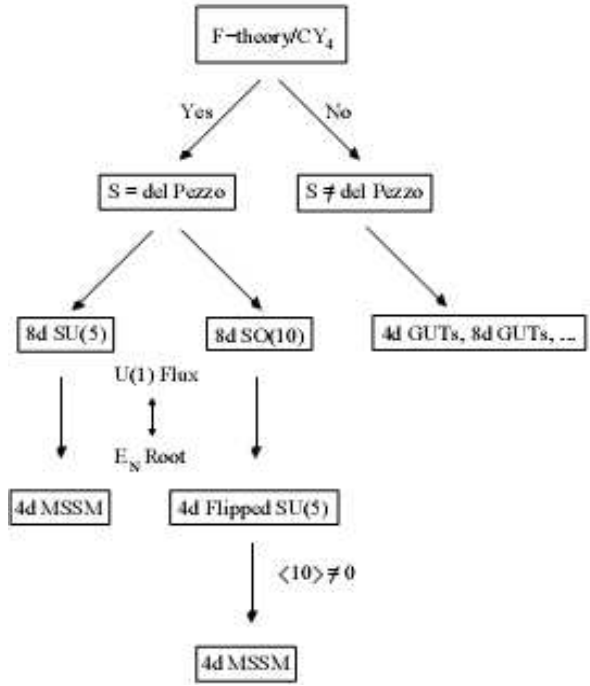}
\caption{\label{fig:2} \small General overview of how GUT breaking constrains the type of GUT model.}
\end{figure}

	We will consider elliptically fibered Calabi-Yau fourfold $X$  with the base $B$ and elliptic fiber $\varepsilon$: 
\begin{center}
\begin{tabular}{ccc}
$\varepsilon$\hspace*{-2mm}&$\rightarrow$&\hspace*{-2mm}$X$\\
&&\hspace*{-2mm}$\downarrow$\\
&&\hspace*{-2mm}$B$
\end{tabular}
\end{center}
This Calabi-Yau fourfold can be  represented by the  Figure 3.
\begin{figure}[htbp]
\centering 
\includegraphics[width=.45\textwidth]{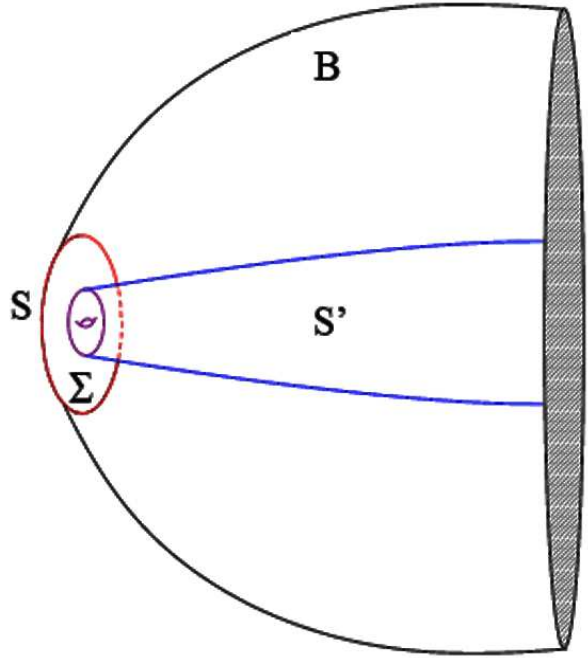}
\caption{\label{fig:3} \small Depiction of F-theory compactified on a local model of a Calabi-Yau fourfold. }
\end{figure}
We shall assume that there exists a Calabi-Yau fourfold which contains the corresponding local enhancement in singularity type. When a del Pezzo surface $S$ intersect $S^{'}$ on a Riemann surface $\Sigma$, the singularity type enhances further. In this case, additional six-dimensional hypermultiplets localize along $\Sigma$.   In terms of four-dimensional superfields, the matter content, localized on a curve $\Sigma$,  consists of chiral superfields.  Schematically this can be represented by the\\ Table 1.
\begin{table}[tbp]
\centering
\begin{tabular}{|clc|}
\hline
Dimension& Space& Ingredient\\
\hline 
10 & $R^{3+1}\times B$ &Gravity\\
8 & $R^{3+1}\times S$ & Gauge Theory\\
6 & $R^{3+1}\times \Sigma$ & Chiral Matter\\
4& $R^{3+1}$& Cubic Interaction Terms\\
\hline
\end{tabular}
\caption{\label{tab:i} Dimensional reduction and matter content of the corresponding space-time.}
\end{table}

	As we can see, we have two points of view, connected with Calabi-Yau fourfolds. One is that Calabi-Yau is the sheaf on $P_{2}$ and as the object in $D(X)$ is stable or unstable in the Teichm$\ddot{u}$ller space of $B+iJ$. The other is that it is the fibered bundle, that can be reduced to four-dimensional theory with the corresponding matter content. After implementation of a higher-dimensional breaking mechanism to obtain four-dimensional models, we can receive the minimal four-dimensional supersymmetric $SU(5)$ Grand Unification Theory with standard Higgs content. 
	
	The moduli space of an open superstring \cite{8.} which is described by $\mbox{Ext}^{i}(Q,Q^{'})$ groups and determined by the diagram \cite{4.} in Figure 4
\begin{figure}[htbp]
\centering 
\includegraphics[width=.25\textwidth]{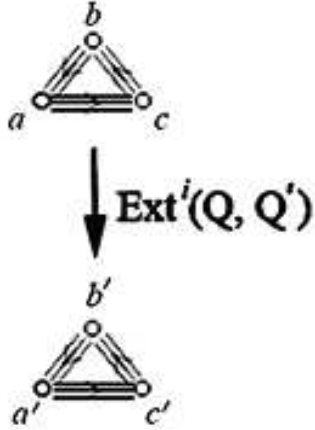}
\caption{\label{fig:4} \small Open superstring that is described by $\mbox{Ext}^{i}(Q,Q^{'})$ group.}
\end{figure}
has the form
\[\hspace*{4.3cm}\begin{array}{ccc}
\mbox{Ext}^{0}(Q,Q^{'})&=&{\newfont{\blackboard}{msbm10 scaled\magstep2}
\newcommand{\Ce}{\mbox{\blackboard\symbol{'103}}}\Ce}^{\hspace*{1mm} aa^{'}+bb^{'}+cc^{'}}\ \ \ \ ,  \\
\mbox{Ext}^{1}(Q,Q^{'})&=&{\newfont{\blackboard}{msbm10 scaled\magstep2}
\newcommand{\Ce}{\mbox{\blackboard\symbol{'103}}}\Ce}^{\hspace*{1mm} 3ab^{'}+3bc^{'}+3ca^{'}}\ .
\end{array}\hspace*{4.5cm} (3)\]
        Substituting in (3) orbifold charges
\[a=b=c=a^{\prime}=b^{\prime}=c^{\prime}=4\]
and using the Langlands hypothesis \cite{9.}, we obtain the realization of (3)
in terms of $SU(5)$ multiplets
\[3\times(24+5_H+\overline{5}_H+5_M+\overline{5}_M+10_M+\overline{10}_M)\ ,\]
where $5_H$ and $\overline{5}_H$  are Higgs multiplets, $\overline{5}^{(i)}_M$ and $10^{(j)}_M$ are multiplets of quark and lepton superpartners.

	As the transition in the Teichm$\ddot{u}$ller space is connected with stability of  the object and this stability is characterized by the moduli space of an open superstring connected with Higgs multiplets, we can see how a higher-dimensional breaking mechanism is connected or associated with four-dimensional GUT Higgs multiplets and symmetry breaking higgs mechanism.

\end{document}